\documentclass[runningheads]{llncs}
\usepackage{graphicx}
\usepackage{color}
\usepackage{amssymb}
\usepackage{amsmath}
\usepackage{graphicx}
\usepackage{caption}
\usepackage{subcaption}
\usepackage{textcomp}
\graphicspath{ {./img/} }
\newcommand{\set}[1]{\mathcal{#1}}
\renewcommand{\vec}[1]{\mathbf{#1}}
\newcommand{\mat}[1]{\mathbf{#1}}
\newcommand{\R}[0]{\mathbb{R}}

\begin{document}
\title{Graph convolutional regression of cardiac depolarization from 
sparse endocardial maps}
\titlerunning{Graph convolutional regression of cardiac depolarization}
\author{Felix Meister\inst{1,3} \and
Tiziano Passerini \inst{2} \and
Chlo\'{e} Audigier \inst{3} \and
\`{E}ric Lluch \inst{3} \and
Viorel Mihalef \inst{2} \and
Hiroshi Ashikaga \inst{4} \and
Andreas Maier \inst{1} \and
Henry Halperin \inst{4} \and
Tommaso Mansi \inst{2}}

\authorrunning{F. Meister et al.}
\institute{Friedrich-Alexander University, Pattern Recognition Lab, Erlangen, 
Germany \and
Siemens Healthineers, Digital Technology \& Innovation, Princeton, USA \and
Siemens Healthineers, Digital Technology \& Innovation, Erlangen, Germany \and
Johns Hopkins University School of Medicine, Cardiac Arrhythmia Service, 
Baltimore, USA}
\maketitle              % typeset the header of the contribution
\begin{abstract}
Electroanatomic mapping as routinely acquired in ablation therapy of
ventricular tachycardia is the gold standard method to identify the arrhythmogenic
substrate.
To reduce the acquisition time and still provide maps with high spatial
resolution, we propose a novel deep learning method based on 
graph convolutional neural networks to estimate the depolarization time in the
myocardium, given sparse catheter data on the left ventricular endocardium,
ECG, and magnetic resonance images.
The training set consists of data produced by a computational
model of cardiac electrophysiology on a large cohort of synthetically
generated geometries of ischemic hearts.
The predicted depolarization pattern has good agreement with
activation times computed by the cardiac electrophysiology model in a
validation set of five swine heart geometries
with complex scar and border zone morphologies. 
The mean absolute error hereby measures
8 ms on the entire myocardium when providing 50\% of the endocardial ground truth
in over 500 computed depolarization patterns. 
Furthermore, when considering a complete animal data set with high
density electroanatomic mapping data as reference, the neural network can 
accurately reproduce the endocardial depolarization pattern, even when a small
percentage of measurements are provided as input features
(mean absolute error of 7 ms with 50\% of input samples). The results show that
the proposed method, trained on synthetically generated data, 
may generalize to real data.

\end{abstract}

%%%%%%%%%%%%%%%%%%%%%%
% Introduction
%%%%%%%%%%%%%%%%%%%%%%

\section{Introduction}
Each year, up to 1 per 1,000 North Americans die of sudden cardiac death,
among which up to 50\% as a result of ventricular tachycardia (VT) \cite{john2012:VAA}.
A well-established therapy option 
for these arrythmias is radiofrequency ablation. While success rates between 80-90\% have been reported
for idiopathic VTs, extinction of recurrent VTs in patients with structural heart disease is only successful in
about 50\% of the cases and requires additional interventions in roughly 30\% of the other half \cite{john2012:VAA}. 
Crucial to ablation success is the identification
of scar regions prone to generate electrical wave re-entry, which is
conventionally realized through electroanatomic mapping
\cite{john2012:VAA}. This technique poses practical challenges to
achieve high spatial resolution and precise localization of the substrate
 \cite{josephson_substrate_2015}. Moreover, it only
provides information about the surface potentials, failing to identify the
complex three-dimensional slow conductive pathways within the myocardium
\cite{ashikaga2007:MRB}.
Imaging (MRI or computed tomography) has great potential in helping define the 
geometry of the substrate \cite{dickfeld2011:MRI,zhang2017multicontrast}, however electrophysiology assessment of
the substrate may not be possible purely based on imaging features.

The use of computational models of cardiac electrophysiology has been explored as a way to combine  imaging and catheter mapping information, to accurately estimate the substrate physical properties and potentially use the resulting model to support clinical decisions
\cite{chinchapatnam_model-based_2008,Dhamala_adaptive_2017,%
pheiffer2017estimation,corrado2018work}.
In this work we propose a novel deep-learning based method to enhance sparse 
left endocardial activation time maps by extrapolating the measurements 
through the biventricular anatomy. A graph convolutional neural 
network is trained to regress the local activation times over the left and 
right ventricle given sparse catheter data acquired on the left ventricular (LV) endocardium,
a surface electrocardiogram (ECG), and 
magnetic resonance (MR) images. The network is trained on data 
produced by a computational model of cardiac electrophysiology on a
large cohort of synthetically generated geometries of ischemic hearts.
Biventricular heart anatomies are sampled from 
a statistical shape model built from porcine imaging data. We evaluate 
the proposed method by applying it to unseen porcine cases with complex scar 
morphology. We further illustrate the potential of the method
to recover endocardial activation from a reduced set of measurements using one unseen 
porcine case with high-resolution LV contact maps.

%%%%%%%%%%%%%%%%%%%%%%
% Method
%%%%%%%%%%%%%%%%%%%%%%
\section{Method}
\begin{figure}[t]
	\centering
	\includegraphics[width=.94\linewidth]{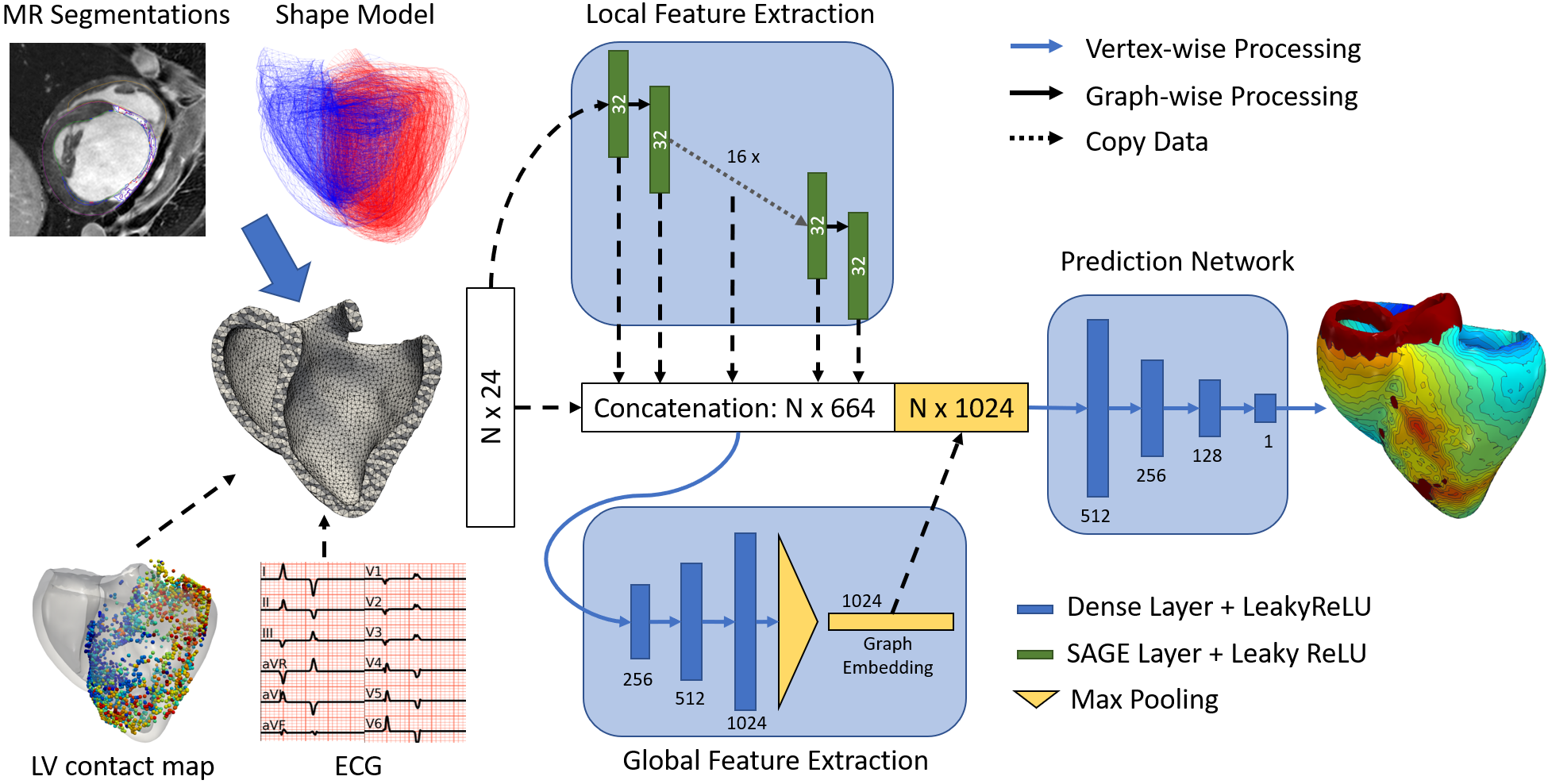}
	\caption{Illustration of the proposed graph convolutional pipeline. ECG information 
 and endocardial measurements are mapped to the vertices of the mesh. 
 Local features are extracted using a cascade of 
 GraphSAGE layers that gather information about the local neighborhood. 
 A global feature vector is computed using a series of fully connected layers and
 max pooling taking the concatenated output of each GraphSAGE layer as input.
 Local activation times (LAT) are regressed using both local features and the global feature 
 vector, and a series of fully connected layers.
}
	\label{fig:network}
\end{figure}

% The key idea is to propagate the sparse left endocardial 
% measurements throughout the biventricular anatomy to obtain local activation 
% times. 
We represent the biventricular heart, segmented from 
medical images, as a tetrahedral mesh \cite{kayvanpour2015:TPC}. The mesh is an undirected 
graph $\set{G} = (\set{V}, \set{E}, \mat{X})$ comprising a set of N vertices $
\set{V}$ and a set of M edges $\set{E}$ connecting pairs of vertices $i$ 
and $j$. Trainable graph convolutional filters with shared weights are applied to all vertices $\in \set{V}$, 
to learn a model of local activation time as a function of vertex-wise features $\mat{X} \in \R^{N \times D}$. 
\\
\textbf{\textit{Graph Convolutional Layer Definition}}
In this work GraphSAGE layers with mean aggregation were chosen \cite{hamilton2017:IRL}. 
Each layer $l$ processes a vector $h_{j}^{l}$, representing any vertex $v_j \in \set{V}$. 
$h_{j}^{0} \in \mat{X}$ denotes the vertex input features. 
GraphSAGE first aggregates information over a vertex $i$'s immediate neighborhood $\set{N}(i)$ comprising all vertices 
$j$ that are connected to $i$ via an edge $e_{ij} \in \set{E}$, 
i.e. $h_{\set{N}(i)}^{(l+1)} = \textrm{mean}({h_{j}^{l}}, \forall j \in \set{N}(i))$.
Using learnable weights $\mat{W}_{\set{N}(i)}$ and $\mat{W}_i$ as well as biases $b_{\set{N}(i)}$ and $b_{i}$,
the layer output $h_{i}^{(l+1)}$ is computed from $h_{\set{N}(i)}^{(l+1)}$ and $h_{j}^{l}$ according to 
$h_{i}^{(l+1)} = \sigma ( \mat{W}_i \cdot h_{i}^{l} + b_{i} + \mat{W}_{\set{N}(i)} \cdot h_{\set{N}(i)}^{(l+1)} + b_{\set{N}(i)
})$
\\
\textbf{\textit{Graph Convolutional Regression of Local Activation Times}}
The feature matrix $\mat{X}$ contains geometric features, ECG information, and
the measured local activation time (LAT) on the LV endocardial surface.
The latter is set to -1 for vertices that are not
associated to measurement points. ECG information consists of QRS interval,
QRS axis, and 12 features representing the positive area under the curve of
the QRS complex in each of the 12 traces. Since each graph convolutional layer 
is applied to all vertices with shared weights, the same ECG features are appended to each vertex.
Geometric features characterize the position of each point in the biventricular
heart. For each tetrahedral mesh, we define a local reference system by three
orthogonal axes:
the left ventricular long axis, the axis connecting the barycenters of the
mitral and tricuspid valves, and an axis normal to both. The apex is chosen 
as the origin of the coordinate system. Each mesh point is
then characterized by the cylindrical coordinates radius, angle, and height. In
addition, we utilize [0, 1]-normalized continuous fields to describe the vertex
position between (0 and 1 respectively) apex and base, endocardium and epicardium, and left ventricle
and right ventricle. Finally we use mutually exclusive categorical features to denote whether a vertex 
belongs to the LV/RV endocardial surface or epicardial surface. We also use another set
of mutually exclusive categorical features for scar, border zone or healthy tissue region. The numerical features are 
[0, 1]-normalized using the bounds computed from the training data set. The LAT
values on the LV endocardium are also normalized as described later.
\\
\textbf{\textit{Network architecture}}
Our network architecture (see Fig.~\ref{fig:network}) is an adaptation of 
PointNet
\cite{qi2017:PDL}. The local feature extraction step using spatial transformer networks is 
replaced by a series of GraphSAGE layers, each leveraging information of the 1-hop neighborhood.
The output of each layer is concatenated to form the vertex-wise local features. 
To obtain a global feature vector, the local features are further processed using multiple 
fully connected layers with shared weights and max pooling. Given local and 
global features the local activation times (LATs) are predicted for each vertex using 
again a series of fully connected layers.
\\
\textbf{\textit{Training Procedure and Implementation}}
The neural network is trained to optimize a loss function $\set{L}= \set{L}_{LAT} + \set{L}_{qrs}$. $\set{L}_{LAT} = 
\frac{1}{N} \sum_{i} \alpha_{i} \|y_i - \hat{y}_i\|^{2}$ is a mean-squared error loss, weighted by $\alpha_{i}$, on 
the predicted LATs $y_i$ and the corresponding ground truth $
\hat{y}_i$. Since there is no guarantee to perfectly match the information provided at the measurement locations, we set $\alpha_{i} = 2$ for the vertices with measurements and $\alpha_{i} = 1$ for all other vertices to guide the network into retaining the measurement data. To preserve QRS duration in the predicted activation pattern, an additional regularization term $\set{L}_{qrs} = \|\textrm{QRSd} - \hat{\textrm{QRSd}}\|^{2}$ is introduced. We approximate the QRS duration 
by computing the difference between maximum and minimum activation time in the 
prediction ($\textrm{QRSd}$) and the ground truth ($\hat{\textrm{QRSd}}$), 
respectively. 

The proposed network is implemented using PyTorch and the Deep Graph Library 
\cite{pytorch,dgl}. Hyperparameters of the network and the optimizer 
were selected using a grid search on a subset of the training data. In particular, 
we choose 20 GraphSAGE layers for the local feature 
extraction, three fully connected layers of size 256, 512, and 1024 for the 
global feature extraction, and four layers of size 512, 256, 128, and 1 for 
the final prediction. For all layers leaky rectified linear units are chosen 
for the activation function. The network is trained for 500 epochs using the 
Adam optimizer \cite{kingma2014adam} with default parameters and an initial learning rate of 5$\times$10$^{-4}$. 
Step-wise learning rate decay is applied every 25 epochs using a decay factor 
of 0.8. No improvement in the validation loss was seen beyond
500 epochs and the network associated with the epoch of minimal validation loss was used  
for evaluation.

%%%%%%%%%%%%%%%%%%%%%%
% Experiments
%%%%%%%%%%%%%%%%%%%%%%
\section{Experiments}
\begin{figure}[t]
\centering
     \begin{subfigure}[t]{0.3\textwidth}
         \centering
         \includegraphics[width=.9\textwidth]{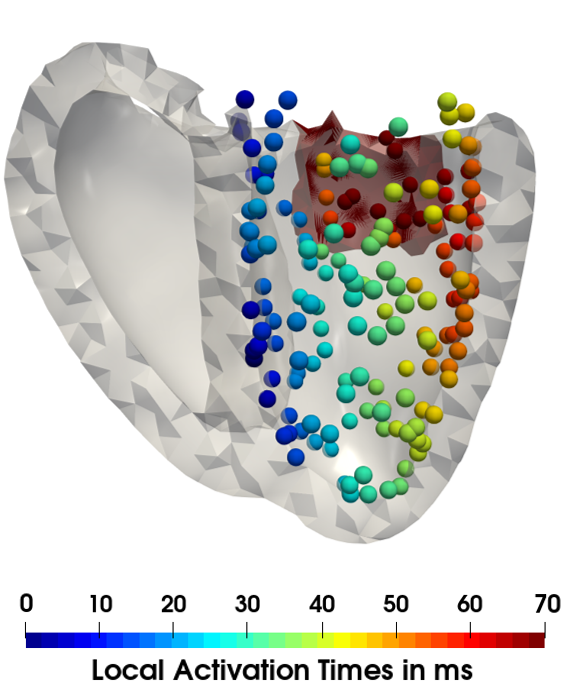}
         \subcaption{}
         \label{fig:comp_aha}
     \end{subfigure}
     \begin{subfigure}[t]{0.3\textwidth}
         \centering
         \includegraphics[width=.9\textwidth]{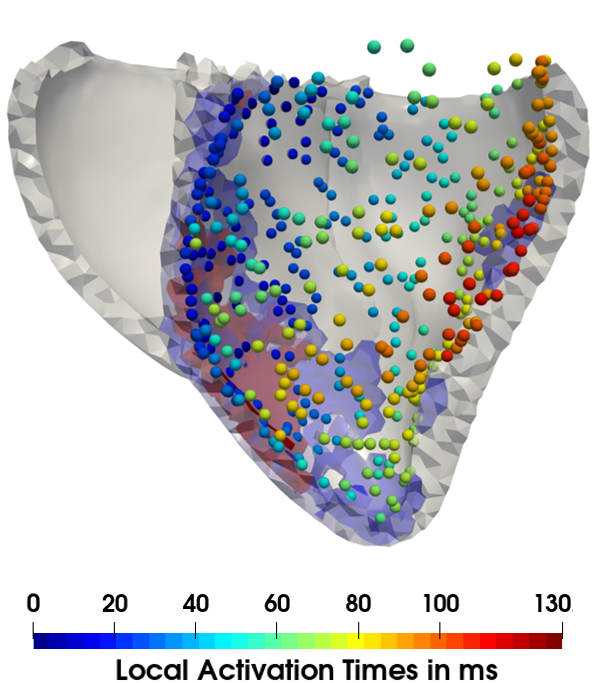}
         \subcaption{}
         \label{fig:comp_scar}
     \end{subfigure}
     \begin{subfigure}[t]{0.3\textwidth}
         \centering
         \includegraphics[width=.9\textwidth]{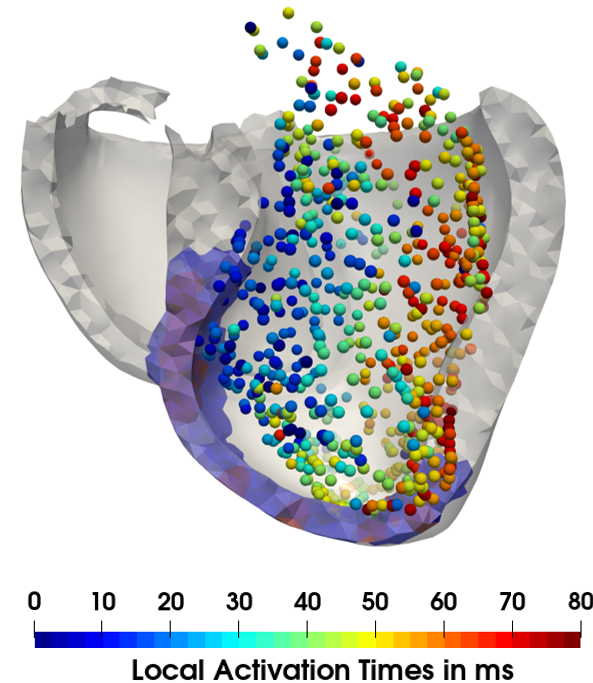}
         \caption{}
         \label{fig:comp_real}
     \end{subfigure}
    \caption{Three examples for geometry and endocardial measurements for (a) synthetic database with simple uniform region randomly set to be scar or border zone; (b) synthetic database with complex scar derived from images; (c) real dataset with complex scar derived from images and high-resolution contact map. In (a) and (b) 25\% of the endocardial ground truth is illustrated.}
    \label{fig:comparison_mesh}
\end{figure}

\begin{figure}[t]
	\centering
	\includegraphics[width=\linewidth]{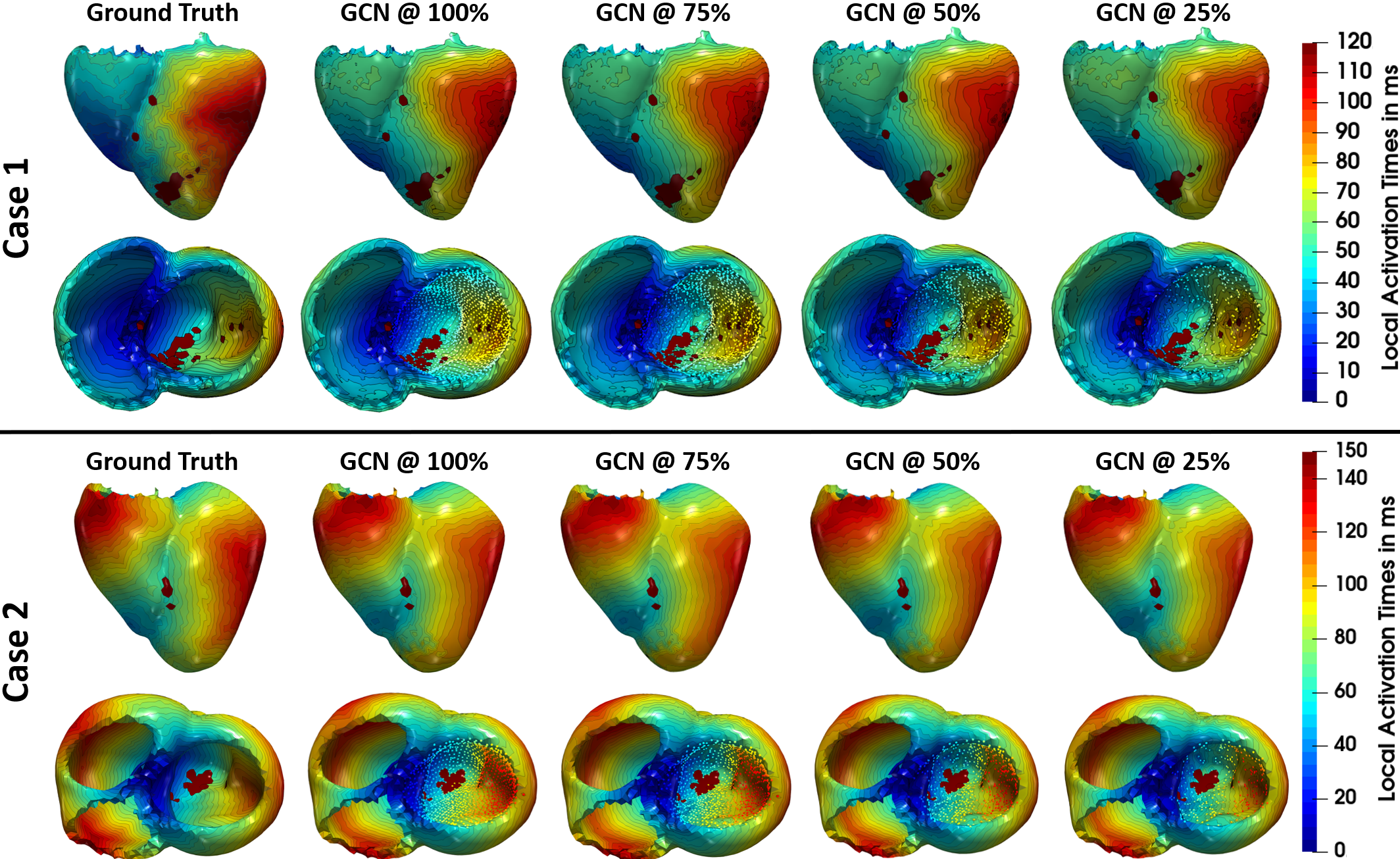}
	\caption{Comparison of prediction results for two cases when providing 25\%, 50\%, 75\%, 
and 100\% of the LV endocardial measurements (colored dots) to the proposed
 network (GCN). The main features of the regressed depolarization pattern
(i.e. wave front curvature and distribution of isochrones) are stable with 
respect to the spatial location and density of the measurements and consistent
with the ground truth.}
	\label{fig:biv_viz}
\end{figure}

\begin{figure}[t]
	\centering
	\includegraphics[width=.7\linewidth]{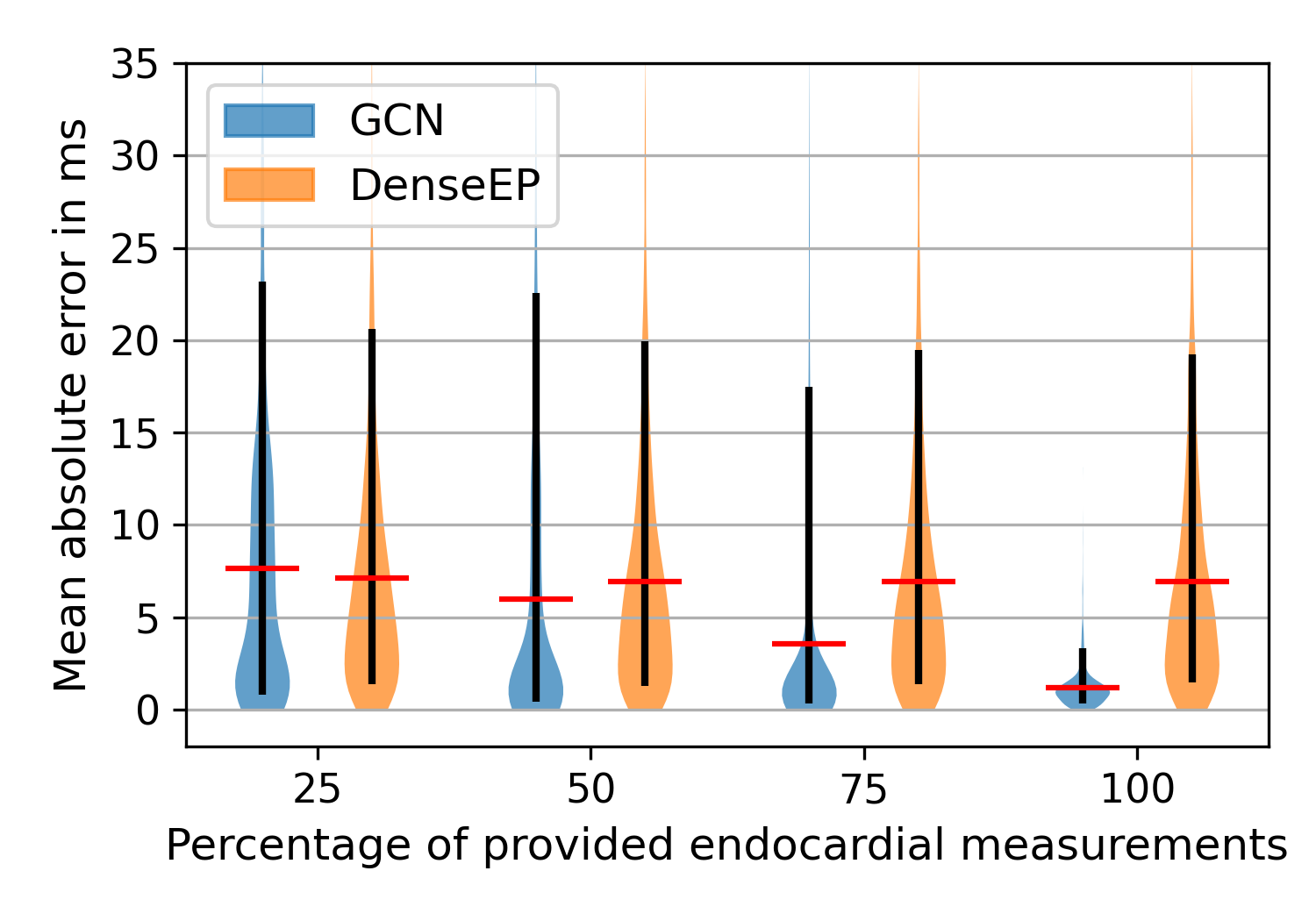}
	\caption{Error statistics on real data experiment. The graph convolutional network (GCN) has a comparable
  performance as the method of reference (DenseEP) for low density LV endocardial measurements
  and outperforms the method of reference as the measurement density increases.}
	\label{fig:err_real_data}
\end{figure}

\begin{figure}[t]
	\centering
	\includegraphics[width=.9\linewidth]{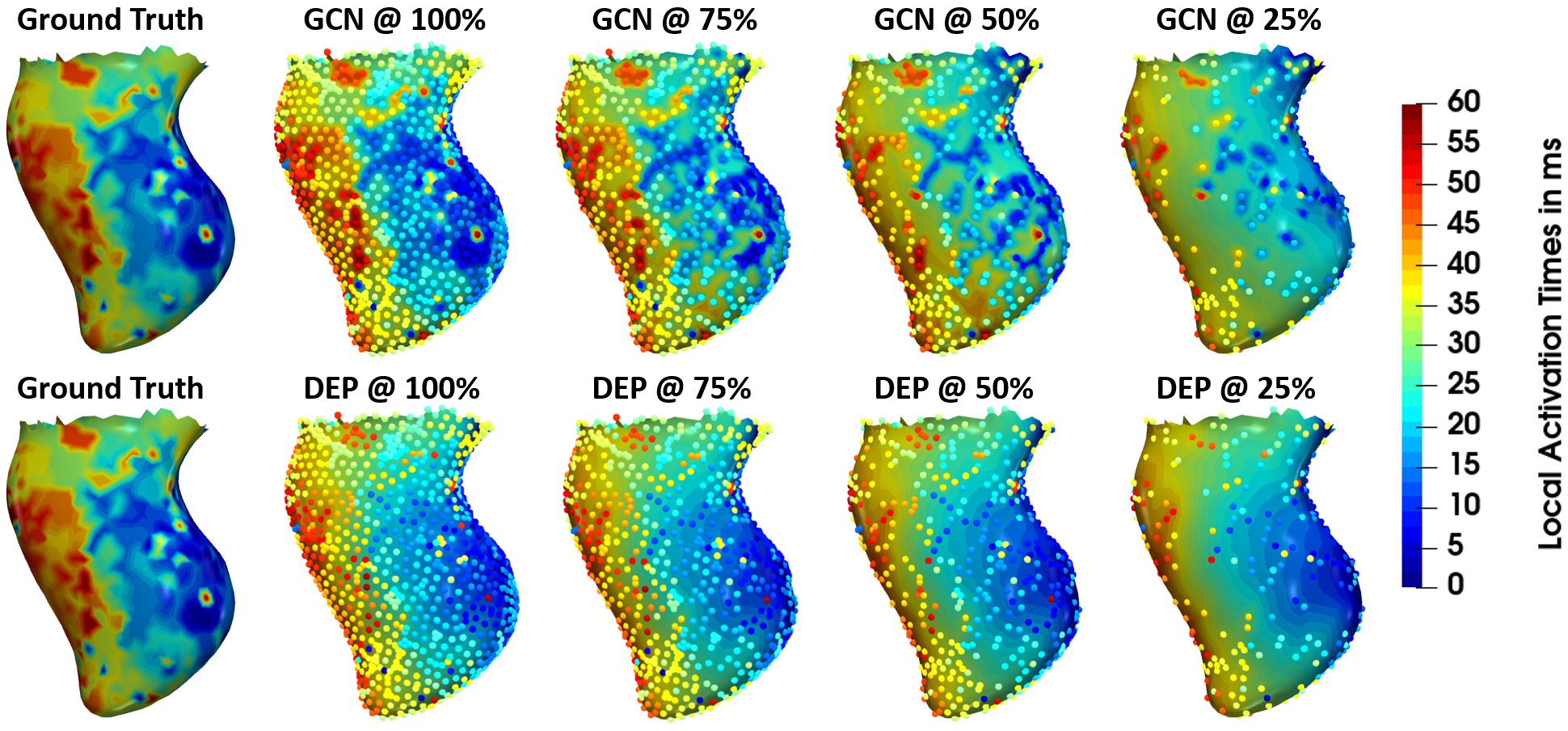}
	\caption{Comparison of prediction results when providing 25\%, 50\%, 75\%, 
and 100\% of the high-resolution catheter measurements (colored dots) to the
proposed network (GCN) and the method of reference (DenseEP). GCN consistently achieves higher 
accuracy in the neighborhood of the measurement points, while DenseEP provides
a smooth and physiologically plausible interpolation of the activation times.}.
	\label{fig:real_data}
\end{figure}

%%%%%%%%%
% Data Generation
%%%%%%%%%
%
\textbf{\textit{Data Generation}}
For training and testing, 16 swine data sets with MR images (MAGNETOM Avanto, Siemens
AG), 12-lead ECG (CardioLab, GE Healthcare), and 
high-resolution left endocardial contact maps (EnSite Velocity System, St. Jude Medical)
 were considered. Additional data sets
were synthetically generated as follows. A statistical shape model
was built from 11 of the 16 swine datasets. A total of 213 swine heart geometries were sampled from it.
In short, heart chamber segmentations were extracted from the medical images 
using a machine learning algorithm \cite{kayvanpour2015:TPC}. The 
segmentations were aligned using point correspondences and rigid registration. 
A mean model was computed over all segmentations and the eigenvectors were computed 
using principal component analysis. The shape model was obtained from the 
linear combinations of the five most informative eigenvectors, which
captured well the global shape variability such as the variation in
ventricle size and wall thickness. Wall thinning as a result of chronic scar
has not been considered due to the limited amount of reference data.
Biventricular heart anatomical models were generated from the chamber surface
meshes using tetrahedralization, mesh tagging, and a rule-based fiber model
 \cite{kayvanpour2015:TPC}. A generic swine torso with pre-defined ECG lead positions
was first manually aligned to the heart chamber segmentations extracted from
one data set, so as to visually fit the visible part of the torso in the cardiac
MR images. Using this as a reference, a torso model was automatically aligned
to any swine heart anatomical model by means of rigid registration. 
A fast graph-based computational electrophysiology model
\cite{pheiffer2017estimation} was utilized to
compute depolarization patterns of the biventricular heart and local activation
times on the LV endocardial surface.
In essence, the activation time for every vertex was computed by finding the shortest path 
to a set of activation points. The edge cost $w_{ij}$ between two vertices $\vec{v}_{i}$ and $\vec{v}_{j}$
was computed as $w_{ij} = l_{ij}/c_{ij}$, where $c_{ij}$ refers to the edge conduction velocity, 
a linear interpolation of the conduction velocity at vertex $\vec{v}_{i}$ and $\vec{v}_{j}$.
$l_{ij} = \sqrt{(\vec{e}_{ij}^T \mat{D} \vec{e}_{ij})}$, with $\vec{e}_{ij} = \vec{v}_{i} - \vec{v}_{j}$, 
refers to a virtual edge length that considers the local anisotropy tensor $\mat{D}$. Using the fiber direction $\vec{f}_{ij}$ from the rule-based fiber model \cite{kayvanpour2015:TPC},
 the anisotropy tensor was computed as $\mat{D} = (1 - r) \vec{f}_{ij}\vec{f}_{ij}^T + r \mat{I}$
with $r$ denoting the anisotropy ratio, i.e. $r = 1 / 3$, and the identity matrix $\mat{I}$.
50 simulations were computed for each anatomical model. 
For each simulation one region of the biventricular heart was randomly selected
from the standard 17 AHA segment model \cite{AHA2002:SMS} to be either scar or border zone tissue (see Fig.~\ref{fig:comp_aha} for an example).
Conduction velocity was assumed uniform in each of five tissue types: normal 
myocardium ($c_{\text{Myo}}$), left and right Purkinje system ($c_{\text{LV}}$ \& $c_{\text{RV}}$), border zone ($c_{\text{BZ}}$) and scar ($c_{\text{Scar}}$). The Purkinje 
system was assumed to extend sub-endocardially with a thickness of 3 mm, as observed in swine hearts. In each
simulation the conduction velocities were randomly sampled within pre-defined
physiological ranges: $c_{\text{Myo}}$ $\in$ [250, 750]\,mm/s, $c_{\text{LV}}$ and $c_{\text{RV}}$ $\in$ [cMyo, 2,500]\,mm/s, and $c_{\text{BZ}}$ $\in$ [100, $c_{\text{Myo}}$]\,mm/s, 
with the exception of scar tissue to which a conduction velocity of 0\,mm/s was always assigned.
The computed endocardial activation time was used to define the endocardial
measurement feature, and it was normalized by subtracting the time
of earliest ventricular activation of each simulation and dividing by the 
computed QRS duration.
The purely synthetic database comprised 10,650
simulations and was randomly split into 90\% for training, 5\% for 
validation, and 5\% for evaluation (``Simple Scar Test'') ensuring that the same anatomical model was
not included in more than one of these partitions. 
Additional 500 simulations were run using the anatomical models generated for the
five left-out swines: 100 simulations per each anatomical model (``Complex Scar Test''). In this case,
scar and border zone regions were defined based on segmentation of the MR images,
while the set up of the computational model was unchanged (see Fig.~\ref{fig:comp_scar} for an example).
\\
\textbf{\textit{Comparison to personalized computational model}}
As a reference method (denoted by DenseEP in this work), we used the same graph-based electrophysiology model 
with local conduction velocities estimated by the 
personalization scheme described in \cite{pheiffer2017estimation}. 
In a first step, local conduction velocities were initialized from the homogeneous tissue conductivities 
$c_{\text{Myo}}$, $c_{\text{LV}}$, $c_{\text{RV}}$, and $c_{\text{BZ}}$, which were
iteratively optimized to match QRS duration and QRS axis. A second step aimed at finding local conductivities 
to minimize the discrepancy between measurements $\hat{y}_i$ and the computed activation times $y_i$.
In this work, a sum of squared distances $\set{L}_{\text{SSD}} = \sum_{i}(\hat{y}_i - y_i)^{2}$ 
on the $N$ vertices where measurements were available was used. 
To find a set of edge weights that minimizes $\set{L}_{\text{SSD}}$ an algorithm inspired by 
neural networks was employed. Hereby, the tetrahedral mesh on which the electrical wave propagation is computed can be seen as a graph that arranges in layers.
The activation time at the vertices where measurements are available (output layer)
 depends on the activation times at the activation points (input layer) 
and on the path followed by the electrical wave. 
The wave propagates to a first of set of vertices (first hidden layer), 
that are connected to the input layer; and recursively propagates 
to other sets of vertices (hidden layers), each connected to the previous.
 Only the paths (sets of edges) connecting vertices in the output layer 
with activation points in the input layer are considered in the optimization step.
\\
For a given path, starting at an activation point $j$ and ending at a measurement point $i$, the activation time in the end point is computed as $y_{i} = t_{j} + \sum_k w_k$ with $t_{j}$ being the initial activation time in $j$, and $\{w_k\}_{1,\ldots,N}$ being the set of N edges along the path. We seek to find the optimal set of edge weights minimizing $\set{L}_{\text{SSD}}$. The weights are iteratively updated by a gradient descent step $w_k^{t + 1} = w_k^{t} - \gamma g$ with the iteration number $t$, the step size $\gamma$, 
and the gradient $g = \frac{\partial \set{L}_{\text{SSD}}}{\partial w_k}$. It follows that $\frac{\partial \set{L}_{\text{SSD}}}{\partial w_{k}} = \frac{\partial \set{L}_{\text{SSD}}}{\partial y_{i}} \frac{\partial y_{i}}{\partial w_{k}}$ with $\frac{\partial \set{L}_{\text{SSD}}}{\partial y_{i}} = -2 (\hat{y}_i - y_i )$ and $\frac{\partial y_{i}}{\partial w_{k}} = 1$. Since an edge may be traversed multiple times, gradients are accumulated on the edges.
The reader is referred to \cite{pheiffer2017estimation} for more details.
%
%
%%%%%%%%%
% Synthetic results
%%%%%%%%%
\begin{figure}[t!]
     \centering
     \begin{subfigure}[b]{0.48\textwidth}
         \centering
         \includegraphics[width=.95\textwidth,trim={0 10pt 0 10pt},clip]{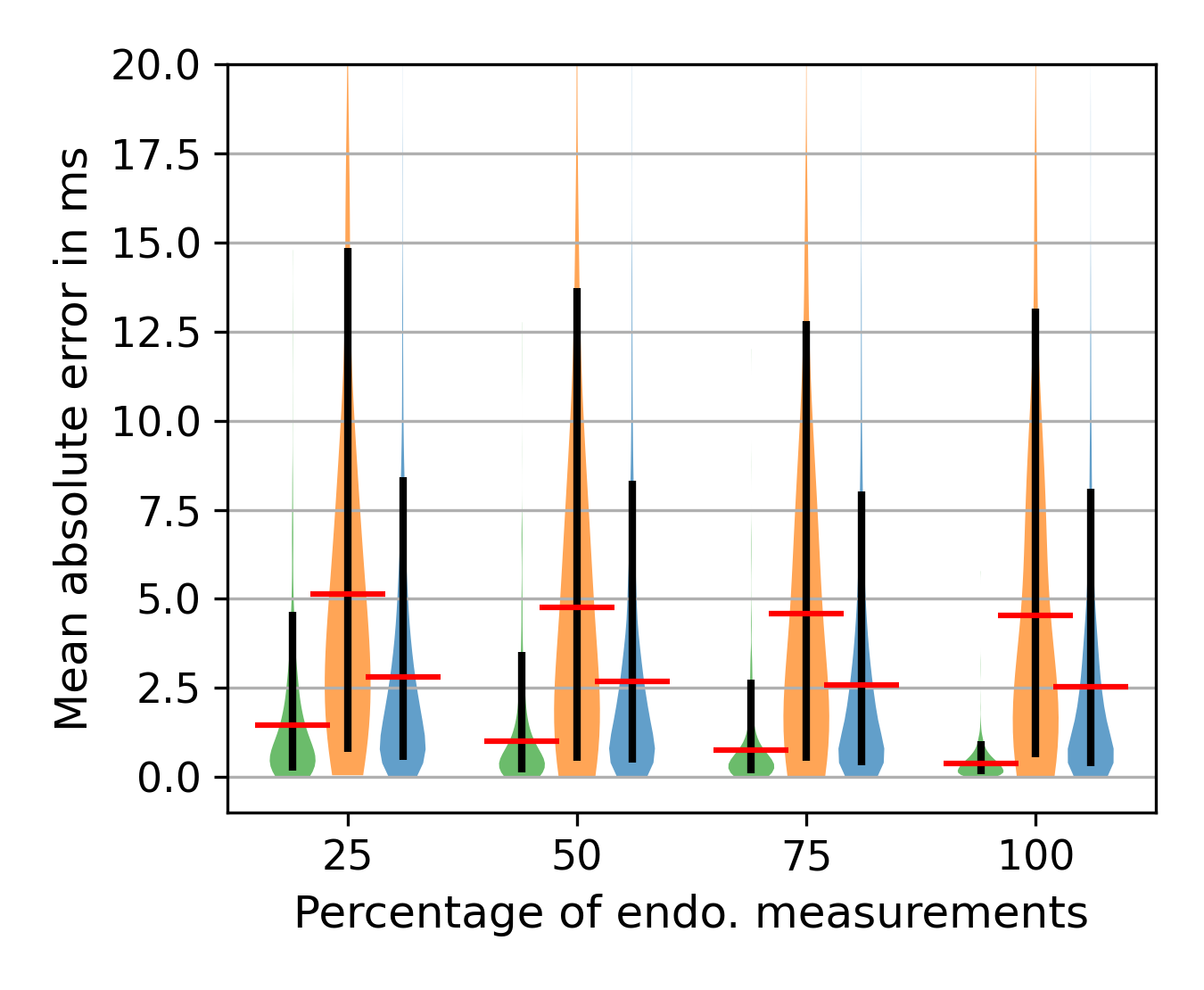}
         \subcaption{Simple Scar (GCN)}
         \label{fig:err_synth}
     \end{subfigure}
     \\
     \begin{subfigure}[t]{0.48\textwidth}
         \centering
         \includegraphics[width=.95\textwidth,trim={0 10pt 0 10pt},clip]{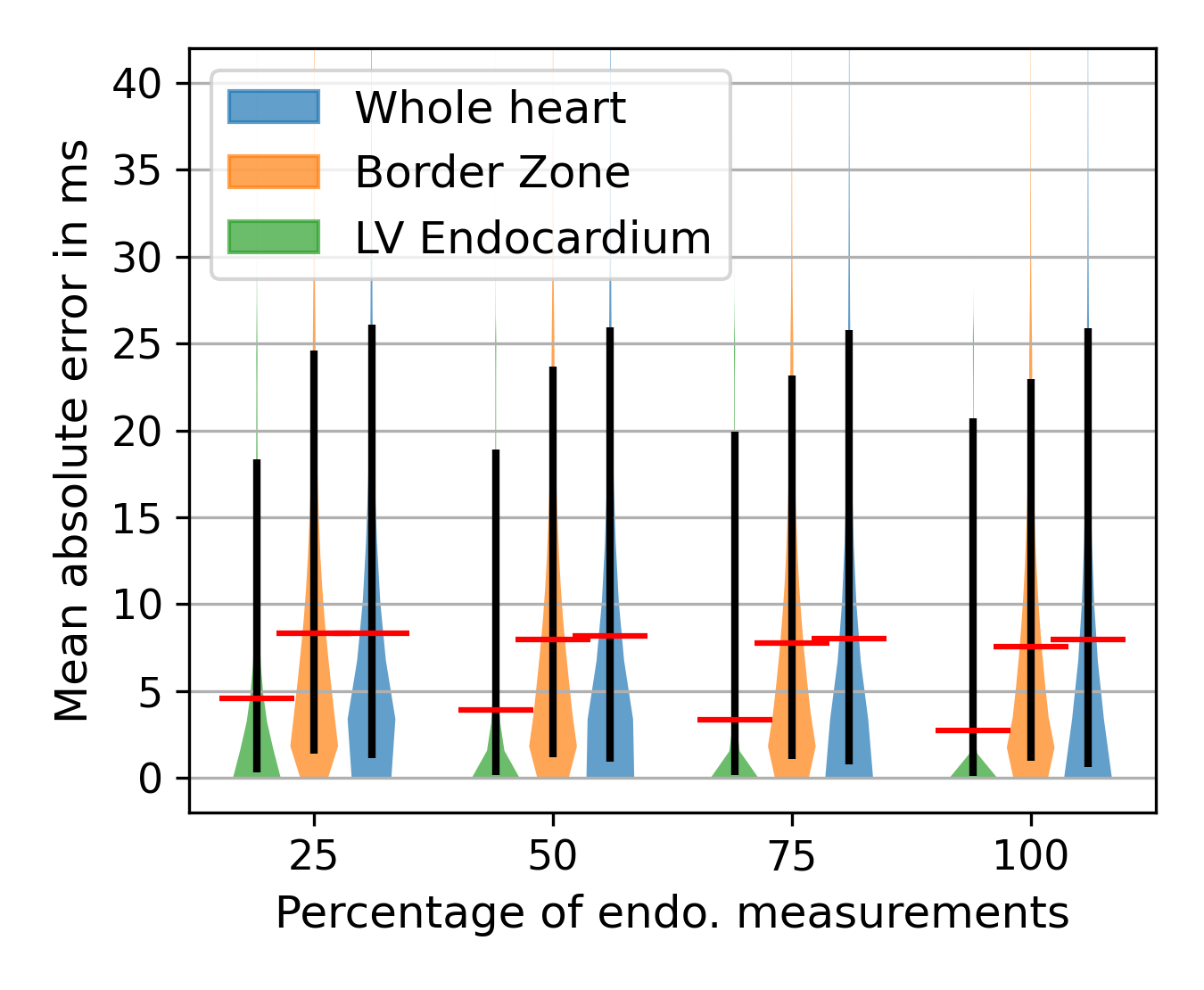}
         \subcaption{Complex Scar (GCN)}
         \label{fig:err_synth_realscar}
     \end{subfigure}
     \begin{subfigure}[t]{0.48\textwidth}
         \centering
         \includegraphics[width=.95\textwidth,trim={0 10pt 0 10pt},clip]{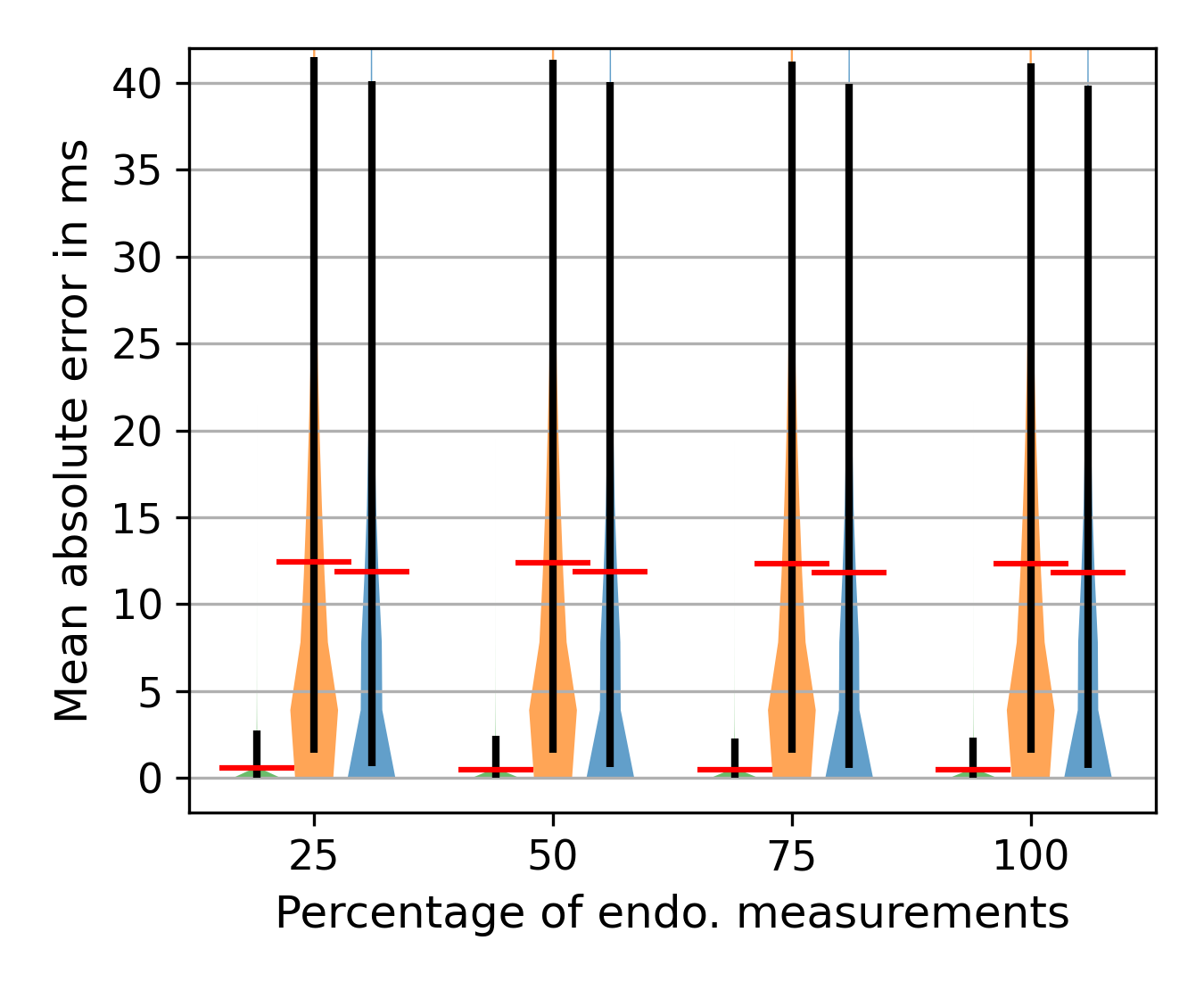}
         \caption{Complex Scar (DenseEP)}
         \label{fig:err_synth_realscar_dep}
     \end{subfigure}
    \caption{Mean absolute error (red: mean, black: 15 to 95 percentile) in 
    local activation time between neural network (GCN) predictions and the ground
    truth values generated by the computational electrophysiology model for Simple Scar Test (a) comprising
    simulations with AHA-based scar or borderzone; and Complex Scar Test (b) with chamber geometries
    and complex scar morphologies not represented by the training database. Good performance
    is achieved on both test sets. As a reference we report in (c) the same 
    metrics on Complex Scar Test for the computational model with densely estimated conductivities (DenseEP).}
    \label{fig:three graphs}
\end{figure}
\\
\textbf{\textit{Evaluation on Synthetic Database}}
The proposed network was trained once on the synthetic training set. We first evaluated the performance of the proposed network on the 5\% left out cases
from the synthetic database, denoted as ``Simple Scar Test'' in the figures. We computed the mean absolute error (L1-error) in
the predicted vertex-wise LAT for (a) the entire biventricular domain, (b) the 
LV endocardial surface, (c) all healthy tissue, and (d) border zone tissue.
For each case we observed the error variation when the endocardial measurements
feature is provided to a decreasing number of vertices (100\%, 75\%, 50\%, and
25\% of the total). The results as seen in 
Fig.~\ref{fig:err_synth} indicate good agreement between the prediction and the
 ground truth even when only using 25\% of the measurements. 
The regressed activation time shows very good agreement with the computed 
values across a wide range of different anatomical morphologies and tissue
physical parameters, and the error remains low when evaluated on the entire
volume of the biventricular heart, on the LV endocardial surface, on
sub-regions of healthy or diseased tissue.
In addition, the mean absolute difference in QRS duration
is less than 1 ms regardless of the endocardial sampling. 
Relative errors w.r.t. the QRS duration measure approximately 2.5\%.
We further evaluated the prediction performance on the testing set consisting of
the 500 additional simulations generated from the anatomical models of the five
pigs not used to generate the statistical shape model (referred to as ``Complex Scar Test''). In this case the
geometry of the heart as well as the complex image-based morphology of scar and
border zone
were not represented by the relatively simple synthetic training database.
Results presented in Fig.~\ref{fig:err_synth_realscar} and Fig.~\ref{fig:biv_viz}
show overall promising
generalization properties of the network, with a slight
decrease in performance compared to the test on fully synthetic data.
Relative errors w.r.t. the QRS duration measure approximately 7\%.
Predictions on the LV endocardial surface were comparatively more accurate than on 
other regions, suggesting the ability of the network to properly
incorporate endocardial measurements even from unseen depolarization patterns. 
However, the accuracy of the prediction 
decreases slightly compared to the case in which we consider synthetically
generated test cases. We believe this to be due to the lack of training samples
showing the same type of patterns in the input features that would be seen
on a graph representing a biventricular heart with a real ischemic scar.
We report in Fig.~\ref{fig:err_synth_realscar_dep} the errors in activation time 
computed by DenseEP on the ``Complex Scar Test''. The results suggest that GCN is less accurate 
than DenseEP at the endocardium. This is consistent with the hypothesis 
that a richer training database (in particular for scar and border zone morphology) 
could help GCN's performance.
%%%%%%%%%
% Real data results
%%%%%%%%%
%
\\
\textbf{\textit{Real Data Results}}
To demonstrate the ability to correctly regress local 
activation times in clinically relevant scenarios, we applied the network, trained on
the synthetic database, to one of the swines from the testing set, which
had a high-density left endocardial electro-anatomical map available.
Catheter measurements included LAT and
peak to peak voltage. The measurement point cloud, which has been curated by an electrophysiologist, 
was spatially registered to the tetrahedral mesh by manual alignment (see Fig.~\ref{fig:comp_real}), 
so that low voltage signals ($<$1.5mV) co-localize with scar and border zone regions. 
Finally, the LAT data was mapped onto the mesh using nearest
 neighbor search. 
 The endocardial measurements were 
randomly subsampled to retain 25\%, 50\%, 75\%, and 100\% of the samples and 
provided as input feature to the network. In all cases, the complete set of 
LV endocardial measurements was used as ground truth and we computed mean absolute
error of vertex-wise LAT on the LV endocardium
(see Fig.~\ref{fig:err_real_data}). The error increases with decreasing number of provided 
endocardial measurements, but remains relatively low and similar to the values 
obtained on unseen geometries with synthetically generated ground truth.
The regressed depolarization pattern is qualitatively correct (see
Fig.~\ref{fig:real_data}) and consistent with the data.
The model identifies the main direction of propagation of the electrical wave,
with early activation on the septal wall and late activation on the free wall.
This qualitative behavior is evident in all solutions provided by the network,
regardless of the amount of subsampling in the endocardial measurements,
suggesting that the extensive training set including physiologically accurate
depolarization patterns produced by a computational model of electrophysiology
provides a strong prior for the prediction. 

For comparison, DenseEP performs consistently well independently of the amount of data
sub-sampling, but with an overall higher average error compared to the network
predictions. The depolarization pattern is consistent
with the data (see Fig.~\ref{fig:real_data}), with a physiologically plausible
smooth interpolation in the measurement gaps owing to the solution
of an inverse problem using the computational electrophysiology model
as a regularizer.
%%%%%%%%%%%%%%%%%%%%%%
% Conclusion
%%%%%%%%%%%%%%%%%%%%%%
\section{Discussion and Conclusion}
Graph convolutional networks are based on a definition of neighborhood
that adapts naturally to the description of complex physical systems. Topology
is one of the main determinants of cardiac electrophysiology, since the 
function of the organ is intrinsically linked to its structure and the 
hierarchy of its components.
The proposed graph convolutional neural network was trained on a large cohort 
of synthetic biventricular depolarization patterns that were generated by a graph-based
computational model of cardiac electrophysiology. Anatomies were sampled
from a statistical shape model. Variability in the depolarization patterns were 
induced by randomizing conduction velocities and randomly adding uniform
regions of scar or border zone on the LV.  Our results were obtained on a left out subset 
from the initial synthetic database as well as on an additional synthetic database 
that incorporated complex scar and border zone distributions derived 
from images. The results show that a graph convolutional network can 
successfully and reliably learn meaningful patterns of activation time as a
function of features that are related to local geometrical and 
functional properties of the heart tissue, when provided with just 
a sparse set of measurements on the LV endocardial surface.
Validation of the method using high density electroanatomic mapping data 
shows that the main features of the regressed depolarization pattern
(i.e. wave front curvature and distribution of isochrones) were stable with 
respect to the spatial location and density of the measurements, which tend to
be incorporated in the depolarization pattern with high accuracy as local
spatial discontinuities in the activation time.
In general, a decrease in the prediction errors was observed the more 
endocardial samples were provided. In case of providing 100\% of the endocardial
samples, however, the network showed to not being able to perfectly match 
the provided information. We hypothesize that the network is not able to
learn to impose the endocardial data points due to the multi-objective loss
function and sharing of weights for processing each vertex. 
Careful construction of a more approriate loss function and the impact
of more complex graph convolutional filters is subject of future work.
While the results suggest that the method generalizes to different meshes 
generated on multiple geometries, the effect of meshing was not studied 
within the scope of this work and is subject to future research.
In addition, the sensitivity of the results on the different 
input features will be analyzed in future work.
Other future directions of this research will focus on
improving the model to regress physiologically plausible activation time
distribution compatible with the available measurements. 
A potential path of improvement comprises the extension of the training
set, for instance by leveraging a more realistic fiber model, e.g. as
presented in \cite{mojica2020:NAO}, or enriching the database 
with complex scar and border zone morphologies.
\\
\textbf{Disclaimer.}
This feature is based on research, and is not commercially available. Due to 
regulatory reasons its future availability cannot be guaranteed.

\bibliographystyle{splncs04}
\bibliography{stacom}

\begin{thebibliography}{10}
\providecommand{\url}[1]{\texttt{#1}}
\providecommand{\urlprefix}{URL }
\providecommand{\doi}[1]{https://doi.org/#1}

\bibitem{ashikaga2007:MRB}
Ashikaga, H., et~al.: Magnetic resonance--based anatomical analysis of
  scar-related ventricular tachycardia: implications for catheter ablation.
  Circulation research  \textbf{101}(9),  939--947 (2007)

\bibitem{chinchapatnam_model-based_2008}
Chinchapatnam, P., et~al.: Model-{Based} {Imaging} of {Cardiac} {Apparent}
  {Conductivity} and {Local} {Conduction} {Velocity} for {Diagnosis} and
  {Planning} of {Therapy}. IEEE Trans. Med. Imaging  \textbf{27}(11),
  1631--1642 (Nov 2008)

\bibitem{corrado2018work}
Corrado, C., et~al.: A work flow to build and validate patient specific left
  atrium electrophysiology models from catheter measurements. Medical image
  analysis  \textbf{47},  153--163 (2018)

\bibitem{Dhamala_adaptive_2017}
{Dhamala}, J., et~al.: Spatially adaptive multi-scale optimization for local
  parameter estimation in cardiac electrophysiology. IEEE T. Med. Imaging
  \textbf{36}(9),  1966--1978 (2017)

\bibitem{dickfeld2011:MRI}
Dickfeld, T., et~al.: Mri-guided ventricular tachycardia ablation: integration
  of late gadolinium-enhanced 3d scar in patients with implantable
  cardioverter-defibrillators. Circ Arrhythm Electrophysiol  \textbf{4}(2),
  172--184 (2011)

\bibitem{hamilton2017:IRL}
Hamilton, W., et~al.: Inductive representation learning on large graphs. In:
  Adv Neural Inf Process Syst. pp. 1024--1034 (2017)

\bibitem{john2012:VAA}
John, R.M., et~al.: Ventricular arrhythmias and sudden cardiac death. The
  Lancet  \textbf{380}(9852),  1520--1529 (2012)

\bibitem{josephson_substrate_2015}
Josephson, M.E., Anter, E.: Substrate {Mapping} for {Ventricular}
  {Tachycardia}. JACC: Clinical Electrophysiology  \textbf{1}(5),  341--352
  (Oct 2015)

\bibitem{kayvanpour2015:TPC}
Kayvanpour, E., et~al.: Towards personalized cardiology: multi-scale modeling
  of the failing heart. PLoS One  \textbf{10}(7) (2015)

\bibitem{kingma2014adam}
Kingma, D.P., Ba, J.: Adam: A method for stochastic optimization. arXiv
  preprint arXiv:1412.6980  (2014)

\bibitem{mojica2020:NAO}
Mojica, M., et~al.: Novel atlas of fiber directions built from ex-vivo
  diffusion tensor images of porcine hearts. Computer Methods and Programs in
  Biomedicine  \textbf{187},  105200 (2020)

\bibitem{AHA2002:SMS}
on~Myocardial~Segmentation, A.H.A.W.G., for Cardiac~Imaging:, R., Cerqueira,
  M.D., Weissman, N.J., Dilsizian, V., Jacobs, A.K., Kaul, S., Laskey, W.K.,
  Pennell, D.J., Rumberger, J.A., Ryan, T., et~al.: Standardized myocardial
  segmentation and nomenclature for tomographic imaging of the heart: a
  statement for healthcare professionals from the cardiac imaging committee of
  the council on clinical cardiology of the american heart association.
  Circulation  \textbf{105}(4),  539--542 (2002)

\bibitem{pytorch}
Paszke, A., et~al.: Pytorch: An imperative style, high-performance deep
  learning library. In: Wallach, H., Larochelle, H., Beygelzimer, A.,
  d\textquotesingle Alch\'{e}-Buc, F., Fox, E., Garnett, R. (eds.) Advances in
  Neural Information Processing Systems 32, pp. 8024--8035. Curran Associates,
  Inc. (2019)

\bibitem{pheiffer2017estimation}
Pheiffer, T., et~al.: Estimation of local conduction velocity from myocardium
  activation time: Application to cardiac resynchronization therapy. In: FIMH.
  pp. 239--248. Springer (2017)

\bibitem{qi2017:PDL}
Qi, C.R., et~al.: Pointnet: Deep learning on point sets for 3d classification
  and segmentation. In: Proceedings CVPR. pp. 652--660 (2017)

\bibitem{dgl}
Wang, M., et~al.: Deep graph library: Towards efficient and scalable deep
  learning on graphs. ICLR Workshop on Representation Learning on Graphs and
  Manifolds  (2019)

\bibitem{zhang2017multicontrast}
Zhang, L., et~al.: Multicontrast reconstruction using compressed sensing with
  low rank and spatially varying edge-preserving constraints for
  high-resolution mr characterization of myocardial infarction. Magnetic
  resonance in medicine  \textbf{78}(2),  598--610 (2017)

\end{thebibliography}
\end{document}